\documentclass[12pt]{article}
\usepackage{graphics}
\usepackage{epsfig}
\usepackage{color}
\usepackage{amsmath}
\usepackage{amssymb}
\def\da{\dot{\alpha}}
\def\db{\dot{\beta}}
\def\dg{\dot{\gamma}}
\def\dd{\dot{\delta}}

\def\L{\Lambda}

\def\a{\alpha}
\def\b{\beta}
\def\g{\gamma}
\def\d{\delta}
\def\e{\varepsilon}
\def\m{\mu}

\def\s{\sigma}
\def\r{\rho}
\def\l{\lambda}
\def\t{\tau}
\def\o{\omega}

\def\vt{\vartheta}
\def\mc{\mathcal}

\def\p{\partial}

\def\ra{\rangle}
\def\dag{\dagger}
\def\wt{\widetilde}
\def\K{\widetilde{K}}
\def\lr{\Leftrightarrow}

\numberwithin{equation}{section} 
\parskip=\medskipamount

\def\Zop{\bbbz}
\def\Rop{\bbbr}

\def\bbbz {{\sf Z\!\!Z}}

\def\bbbr {{\rm I\!R}}

\def\RR{R-R }
\newcommand{\ket}[1]{|#1\rangle}

\begin{document}

\thispagestyle{empty}
\addtocounter{page}{-1}
\def\thefootnote{\fnsymbol{footnote}}
\begin{flushright}
  hep-th/0308062 \\
  AEI-2003-65 \\
  KCL-MTH-2003-14	\\
  SPIN-2003/23 \\
  ITP-2003/36
\end{flushright}

\vskip 0.5cm

\begin{center}\LARGE
{\bf On the Uniqueness of \\ Plane-wave String Field Theory}
\end{center}

\vskip 1.0cm

\begin{center}
{\large A. Pankiewicz\,\footnote{E-mail address: {\tt apankie@aei.mpg.de}}${}^{,**}$ and
B. Stefa\'nski, jr.\,\footnote{E-mail address: {\tt stefansk@phys.uu.nl}}}

\vskip 0.5cm

{\it $^*$ Max-Planck-Institut f\"ur Gravitationsphysik, Albert-Einstein-Institut \\
Am M\"uhlenberg 1, D-14476 Golm, Germany}

\vskip 0.2cm 

{\it $^{**}$ Department of Mathematics, King's College London \\
Strand, London WC2R 2LS, United Kingdom}

\vskip 0.5cm

{\it $^\dagger$ Spinoza Institute, University of Utrecht \\
Postbus 80.195, 3508 TD Utrecht, The Netherlands}
\end{center}

\vskip 1.0cm

\begin{center}
August 2003
\end{center}

\vskip 1.0cm

\begin{abstract}
\noindent We prove that the two interaction Hamiltonians of light-cone 
closed superstring field theory in the plane-wave background present in 
the literature are identical. 
\end{abstract}

\vfill

\setcounter{footnote}{0}
\def\thefootnote{\arabic{footnote}}
\newpage

\renewcommand{\theequation}{\thesection.\arabic{equation}}

\section{Introduction}\label{introduction}
\setcounter{equation}{0}
Following the discovery of the plane-wave solution of Type IIB 
supergravity~\cite{Blau:2001ne}~\footnote{For previous work on supergravity
plane-wave solutions see~\cite{Kowalski-Glikman:wv,Hull:vh}.}, the 
spectrum and 
superalgebra of the 
free superstring theory in this background were found in the light-cone 
gauge~\cite{Metsaev:2001bj,Metsaev:2002re}. The theory possesses a unique groundstate and a 
tower of states with energies proportional to
\begin{equation}
\omega_n=\sqrt{n^2+(\mu\alpha^\prime p^+)^2}\,,
\end{equation}
where $n\in\Zop$ and $\mu$ and $p^+$ are the R-R field-strength and light-cone momentum 
respectively. The plane-wave background has also become 
important because of its interpretation as a Penrose limit~\cite{Blau:2002dy} of the 
$AdS_5\times S^5$ space-time. In this setting, the AdS/CFT correspondence 
has been identified as a relation between string theory in the large $\mu$ 
limit and the ${\cal N}=4$ $U(N)$ SYM gauge theory in a non-'t Hooft limit 
where not only $N$, but also $J$, a chosen U(1) R-charge, is taken to be 
large, with the ratio $J^2/N$ fixed~\cite{Berenstein:2002jq}. A subset of so-called BMN 
operators has been identified in the gauge theory which corresponds to 
string states. These operators have an expansion in terms of an 
effective coupling constant 
$\lambda^\prime=g_{\text{YM}}^2N/J^2=(\mu\alpha^\prime p^+)^{-2}$ and 
effective genus counting parameter 
$g_2^2=(J^2/N)^2=g_s^2(\mu\alpha^\prime p^+)^4$
~\cite{Kristjansen:2002bb,Constable:2002hw}, and the gauge/gravity 
correspondence in this background is given by~\cite{Gross:2002su} 
\begin{equation}
\frac{1}{\mu}H_s=\Delta-J\,,
\end{equation}
viewed as an operator identity between the Hilbert spaces of
string theory and the BMN sector of gauge theory. This correspondence 
has been placed on a firm footing at the level of planar graphs, or 
equivalently at the level of free string 
theory~\cite{Berenstein:2002jq,Kristjansen:2002bb,Constable:2002hw,
Gross:2002su,Santambrogio:2002sb}. 
At the non-planar/string interaction level there is also good evidence 
that, at 
least for so-called impurity preserving amplitudes, the operator identity 
above is valid~\cite{Gross:2002mh,Beisert:2002bb,Constable:2002vq,Gomis:2002wi,Beisert:2002ff,
Pearson:2002zs,Roiban:2002xr,Gomis:2003kj}, see also~\cite{Pankiewicz:2003pg,Plefka:2003nb} for recent reviews. 

An essential ingredient in the understanding of string theory in the 
plane-wave background is the knowledge of string interactions. 
Unfortunately, the background has only been quantized in light-cone gauge 
and so conformal field theory tools such as vertex operators cannot be used here.
\footnote{In flat space it is possible to develop vertex 
operator techniques even in light-cone gauge~\cite{Green:1981xx}. This is 
aided by the 
presence of a classical conformal invariance of the equations of motion in 
light-cone gauge, as well as by the existence of angular momentum 
generators $J^{-I}$.}
The only known way of studying string interactions in the 
plane-wave comes from light-cone string field 
theory~\cite{Mandelstam:1973jk,Mandelstam:1974fq,Mandelstam:1974hk,
Green:1983tc,Green:1983hw,Cremmer:1974jq,Cremmer:1975ej}. 
In this 
formalism the generators of the supersymmetry algebra are divided into two 
sets of operators: the kinematical and the dynamical. The former, such as 
the transverse momenta $P^I$, do not receive corrections in the string 
coupling $g_s$, while the latter, which include the Hamiltonian, 
are modified order by order in the string coupling. For 
example
\begin{equation}
H_s=H_2+g_sH_3+\dots\,,
\end{equation}
where $H_2$ is the free-string Hamiltonian and $H_3$ represents the 
process of one string splitting into two (as well as the time-reversal of 
this interaction). When computing string interactions it is most 
convenient to write $H_3$ as an operator in the three-string Hilbert 
space~\cite{Cremmer:1974jq,Cremmer:1975ej}.

The interaction Hamiltonian $H_3$ is constructed by 
requiring two conditions. Firstly, the process is to be smooth on 
the world-sheet; this is equivalent to demanding the 
supercommutation relations between the interaction Hamiltonian and the 
kinematical generators 
be satisfied. In the operator formalism this is enforced by a coherent 
state of the three-string Hilbert space often denoted by 
$\ket{V}$. Secondly, $H_3$ is required to satisfy the supersymmetry 
algebra relations involving the Hamiltonian and the dynamical 
supercharges at next-to-leading order in the string coupling. These 
conditions require that 
\begin{equation}
\ket{H_3}={\cal P}\ket{V}\,,
\end{equation}
where ${\cal P}$ is the so-called prefactor which, in the oscillator basis, 
is polynomial in the creation operators. 

Originally~\cite{Spradlin:2002ar,Spradlin:2002rv,Pankiewicz:2002gs,Pankiewicz:2002tg}, when $H_3$ was constructed in the plane-wave 
background, the oscillator basis expression was built on the state 
$\ket{0}$ which has energy $4\mu$ and (hence) is not the 
groundstate of the theory.\footnote{For the precise definitions of 
$\ket{0}$ and $\ket{v}$ see section~\ref{section2}.} Rather, it is 
smoothly 
connected to the $SO(8)$ invariant flat space state $\ket{0}_{\m=0}$ on 
which the flat spacetime interaction vertex was built~\cite{Green:1983hw}. 
We will refer to $H_3$ constructed on this state as the 
$SO(8)$ solution throughout this paper  
\begin{equation}
\ket{H_3}_{SO(8)}={\cal P}_{SO(8)}\ket{V}_{SO(8)}\,.
\end{equation}
The presence of the \RR flux in the plane-wave background breaks the transverse $SO(8)$ symmetry of the metric to 
$SO(4)\times SO(4)\times\Zop_2$, where
the discrete $\Zop_2$ is an $SO(8)$ transformation that exchanges the two 
transverse $\Rop^4$ subspaces of the plane-wave.
Based on this $\Zop_2$ symmetry it was argued~\cite{Chu:2002eu}
that one should in fact construct $H_3$ on the true 
groundstate of the theory: $\ket{v}$. A solution of the kinematical 
constraints based on this state was given in~\cite{Chu:2002wj}, while the dynamical 
constraints were solved in~\cite{Pankiewicz:2003kj}; this solution will be called the 
$SO(4)^2$ solution here
\begin{equation}
\ket{H_3}_{SO(4)^2}={\cal P}_{SO(4)^2}\ket{V}_{SO(4)^2}\,.
\end{equation}
The two interaction Hamiltonians appeared to be quite different, and it 
was not, {\em a priori} clear, if they should give the same 
physics.\footnote{Some 
evidence that they were in fact identical was already presented 
in~\cite{Pankiewicz:2003kj}.}

In this paper we prove that the two interaction Hamiltonians are 
identical when viewed as operators acting on the three-string Hilbert 
space. The proof is presented in section~\ref{section2} for the 
supergravity modes only, and generalized in section~\ref{section3} to the 
full three-string Hamiltonian. Two appendices are provided in which 
our conventions are summarized and some of the computational details are 
presented.


\section{The equivalence of the $SO(8)$ and $SO(4)^2$ \\ formalisms in supergravity}\label{section2}

In this section we prove that the supergravity three-string interaction
vertices constructed in the $SO(8)$ formalism in~\cite{Spradlin:2002ar}
and in the $SO(4)^2$ formalism in~\cite{Pankiewicz:2003kj} are
identical to each other. Recall the fermionic part of the light-cone 
action on the plane wave~\cite{Metsaev:2001bj}
\begin{equation}
S_{\mbox{\scriptsize ferm.}(r)}=\frac{1}{8\pi}\int\,d\t\int_0^{2\pi|\a_r|}\,d\s_r
[i(\bar{\vt}_r\dot{\vt}_r+\vt_r\dot{\bar{\vt}}_r)
-\vt_r\vt'_r+\bar{\vt}_r\bar{\vt}'_r-2\m\bar{\vt}_r\Pi\vt_r]\,,
\end{equation}
where $r=1$, $2$, $3$ denotes the $r$th string, $\a_r\equiv \a' p^+_r$ and 
$e(\a_r)\equiv\a_r/|\alpha_r|$. 
$\vt^a_r$ is a complex, positive chirality SO(8) spinor, $\dot{\vt}_r\equiv\p_{\t}\vt_r$, $\vt'_r\equiv\p_{\s_r}\vt_r$ 
and $\Pi_{ab}\equiv(\g^1\g^2\g^3\g^4)_{ab}$ is symmetric, traceless and squares to one.
The mode expansions of $\vt^a_r$ and its conjugate momentum $\l^a_r\equiv~{\bar\theta}_r^a/4\pi$ at $\t=0$ are
\begin{equation}\label{modeex}
\begin{split}
\vt^a_r(\s_r) & =\vt^a_{0(r)}+\sqrt{2}\sum_{n=1}^{\infty}
\bigl(\vt^a_{n(r)}\cos\frac{n\s_r}{|\a_r|}+\vt^a_{-n(r)}\sin\frac{n\s_r}{|\a_r|}\bigr)\,,\\
\l^a_r(\s_r) & =\frac{1}{2\pi|\a_r|}\bigl[\l^a_{0(r)}+\sqrt{2}\sum_{n=1}^{\infty}
\bigl(\l^a_{n(r)}\cos\frac{n\s_r}{|\a_r|}+\l^a_{-n(r)}\sin\frac{n\s_r}{|\a_r|}\bigr)\bigr]\,.
\end{split}
\end{equation}
The Fourier modes satisfy $2\l_{n(r)}^a=|\a_r|\bar{\vt}^a_{n(r)}$
and the canonical anti-commutation relations for the fermionic coordinates yield the anti-commutation rules
\begin{equation}
\{\vt^a_r(\s_r),\l^b_s(\s_s)\}=\d^{ab}\d_{rs}\d(\s_r-\s_s)\qquad\lr\qquad\{\vt^a_{n(r)},\l^b_{m(s)}\}=\d^{ab}\d_{nm}\d_{rs}\,.
\end{equation}
The fermionic normal modes are defined via ($e(0)\equiv 1$)
\begin{equation}\label{normal}
\vt_{n(r)}=\frac{c_{n(r)}}{\sqrt{|\a_r|}}\left[(1+\r_{n(r)}\Pi)b_{n(r)}
+e(\a_r)e(n)(1-\r_{n(r)}\Pi)b_{-n(r)}^{\dag}\right]\,,\qquad n\in\Zop\,,
\end{equation}
and break the $SO(8)$ symmetry to $SO(4)\times SO(4)$. Here
\begin{equation}
\r_{n(r)}=\r_{-n(r)}=\frac{\o_{n(r)}-|n|}{\m\a_r}\,,\qquad
c_{n(r)}=c_{-n(r)}=\frac{1}{\sqrt{1+\r_{n(r)}^2}}\,. 
\end{equation}
These modes satisfy
$\{b^a_{n(r)},b^{b\,\dag}_{m(s)}\}=\d^{ab}\d_{nm}\d_{rs}$. 
The two states $\ket{v}$ and $\ket{0}$, on which the interaction 
Hamiltonians are constructed, are then annihilated by all $b_n(r)$ for 
$n\neq 0$ with
\begin{equation}
\theta_0^a\ket{0}=0\,,\qquad b_0^a\ket{v}=0\,.
\end{equation}
We use a $\g$-matrix representation in which
\begin{equation}
\Pi =
\begin{pmatrix}
\d_{\a_1}^{\b_1}\d_{\a_2}^{\b_2} & 0 \\ 0 & -\d^{\da_1}_{\db_1}\d^{\da_2}_{\db_2}
\end{pmatrix}\,,
\end{equation}
where $\a_k$, $\da_k$ ($\b_k$, $\db_k$) are two-component Weyl indices of $SO(4)_k$.\footnote{See appendix~\ref{appA} 
for our conventions.}
Hence, $(1\pm\Pi)/2$
projects onto the $({\bf 2},{\bf 2})$ and $({\bf 2'},{\bf 2'})$ of $SO(4)\times SO(4)$,
respectively, and
\begin{equation}
\{b_{n(r)\,\a_1\a_2},b^{\b_1\b_2\,\dag}_{m(s)}\}=\d_{\a_1}^{\b_1}\d_{\a_2}^{\b_2}\d_{nm}\d_{rs}\,,\qquad
\{b_{n(r)\,\da_1\da_2},b^{\db_1\db_2\,\dag}_{m(s)}\}=\d_{\da_1}^{\db_1}\d_{\da_2}^{\db_2}\d_{nm}\d_{rs}\,.
\end{equation}
The fermionic contribution to the free string light-cone Hamiltonian is
\begin{equation}
H_{2(r)}=\frac{1}{\a_r}\sum_{n\in\Zop}\o_{n(r)}\bigl(
b^{\a_1\a_2\,\dag}_{n(r)}b_{n(r)\,\a_1\a_2}+b^{\da_1\da_2\,\dag}_{n(r)}b_{n(r)\,\da_1\da_2}\bigr)\,,
\end{equation}
and we have neglected the zero-point energy that is canceled by the 
bosonic contribution. 

\subsection{The kinematical part of the vertex} 

The fermionic contributions to $\ket{V}$ - the kinematical part of the 
supergravity vertices - in the $SO(8)$ and 
$SO(4)^2$ formalisms are respectively ($\b_r\equiv-\frac{\a_r}{\a_3}$
and $\a_1+\a_2+\a_3=0$) 
\begin{align}
\label{so(8)}
|E^0_b\ra_{SO(8)} & = \prod_{a=1}^8\left[\sum_{r=1}^3\l_{0(r)}^a\right]|0\ra_{123}\,,\\
\label{so(4)}
|E^0_b\ra_{SO(4)^2} & = \exp\left(\sum_{r=1}^2\sqrt{\b_r}
\bigl(b^{\a_1\a_2\,\dag}_{0(3)}b^{\dag}_{0(r)\,\a_1\a_2}+b^{\da_1\da_2\,\dag}_{0(3)}b^{\dag}_{0(r)\,\da_1\da_2}
\bigr)
\right)|v\ra_{123}\,.
\end{align}
To relate these two expressions recall that ({\it cf}. equation~\eqref{normal})
\begin{align}
\l_{0(3)}^{\a_1\a_2} & =-\sqrt{-\frac{\a_3}{2}}b^{\a_1\a_2}_{0(3)}\,,\qquad
\l_{0(3)}^{\da_1\da_2}=\sqrt{-\frac{\a_3}{2}}b^{\da_1\da_2\,\dag}_{0(3)}\,,\\
\l_{0(r)}^{\a_1\a_2} &=\sqrt{\frac{\a_r}{2}}b^{\a_1\a_2\,\dag}_{0(r)}\,,\qquad\;\;\;\;
\l_{0(r)}^{\da_1\da_2}=\sqrt{\frac{\a_r}{2}}b^{\da_1\da_2}_{0(r)}\,,
\end{align}
and 
\begin{equation}\label{0-v}
|0\ra_3= -\prod_{\a_1,\,\a_2}b_{0(3)\,\a_1\a_2}^{\dag}|v\ra_3\,,\quad
|0\ra_r= \prod_{\da_1,\,\da_2}b_{0(r)\,\da_1\da_2}^{\dag}|v\ra_r\,.
\end{equation}
The relative sign in~\eqref{0-v} is not fixed and has been chosen for 
convenience. Then it is easy to show that
\begin{equation}\label{kinrel0}
|E^0_b\ra_{SO(8)} = -\left(\frac{\a_3}{2}\right)^4\prod_{\da_1,\,\da_2}
\bigl(\sqrt{\b_1}b_{0(2)}^{\dag}-\sqrt{\b_2}b_{0(1)}^{\dag}\bigr)_{\da_1\da_2}
|E^0_b\ra_{SO(4)^2}\,.
\end{equation}
By construction, both $|E^0_b\ra_{SO(8)}$ and $|E^0_b\ra_{SO(4)^2}$ satisfy the world-sheet continuity conditions.
Hence, the combination 
$\prod\limits_{\da_1,\,\da_2}\bigl(\sqrt{\b_1}b_{0(2)}^{\dag}-\sqrt{\b_2}b_{0(1)}^{\dag}\bigr)_{\da_1\da_2}$ 
has to commute with the kinematical constraints, and so can be re-written in terms of the 
(zero-mode of the) fermionic prefactor constituent $Z_{\da_1\da_2}$ (in 
the notation of~\cite{Pankiewicz:2002gs}).
In fact
\begin{equation}\label{rel_kinematical}
\left(\frac{2}{\a_3}\right)^4(1-4\m\a K)^{2}|E^0_b\ra_{SO(8)} =
-\prod_{\da_1,\,\da_2}Z_{0\,\da_1\da_2}|E^0_b\ra_{SO(4)^2} 
\equiv
\frac{1}{12}Z_0^4|E^0_b\ra_{SO(4)^2}\,.
\end{equation}
The factor of $\left(\frac{2}{\a_3}\right)^4(1-4\m\a K)^{2}$ was introduced in the $SO(8)$ formalism  
as the overall normalization of the cubic vertex. 


\subsection{Prefactor} 
In order to proceed further, we have to re-write the prefactor of the 
$SO(8)$ 
formulation in a manifestly $SO(4)\times SO(4)$ invariant form using 
the $\g$-matrix representation given in appendix~\ref{appA}. The prefactor 
is~\cite{Pankiewicz:2002tg,Spradlin:2002ar}\footnote{When no confusion arises we will suppress the subscript `0' in what follows.}
\begin{equation}
{\mc P}_{SO(8)} = \bigl(K^I\K^J-\frac{\m\a}{\a'}\d^{IJ}\bigr)v_{IJ}(Y)\,.
\end{equation}
Here $K^I$ and $\wt{K}^I$ are the bosonic constituents commuting with the world-sheet continuity conditions (for their explicit 
expressions see e.g.~\cite{Pankiewicz:2002tg})
and $v_{IJ}=w_{IJ}+y_{IJ}$ with\footnote{Compared 
to~\cite{Pankiewicz:2002tg} we have redefined 
$\sqrt{-\frac{\a'}{\a}}Y_{\text{there}}=Y_{\text{here}}$.}
\begin{align}
w^{IJ} & = \d^{IJ}+\frac{1}{4!}t^{IJ}_{abcd}Y^aY^bY^cY^d+\frac{1}{8!}\d^{IJ}\e_{abcdefgh}Y^a\cdots Y^h\,,\\
y^{IJ} & = -\frac{i}{2!}\g^{IJ}_{ab}Y^aY^b-\frac{i}{2\cdot 6!}\g^{IJ}_{ab}{\e^{ab}}_{cdefgh}Y^c\cdots Y^h\,,
\end{align}
and $t^{IJ}_{abcd}=\g^{IK}_{[ab}\g^{JK}_{cd]}$. The positive and negative chirality parts of $Y^a$ are\footnote{Here the chirality 
refers to either of the two $SO(4)$'s.}
\begin{align}
Y^{\a_1\a_2} & = \sum_{r=1}^3\sum_{n\ge 0}\bar{G}_{n(r)}b_{n(r)}^{\dag\,\a_1\a_2}\,,\\
Y^{\da_1\da_2} & = -(1-4\m\a K)^{-1/2}\sum_{r,s=1}^2\e^{rs}\sqrt{\b_s}b_{0(r)}^{\da_1\da_2} 
+\sum_{r=1}^3\sum_{n>0}U_{n(r)}\bar{G}_{n(r)}b_{n(r)}^{\dag\,\da_1\da_2}\,,
\end{align}
where $\bar{G}$ is defined in~\cite{Pankiewicz:2003kj}.
Note in particular that the zero-mode of $Y^{\da_1\da_2}$ is an annihilation operator. 
If we want to suppress the spinor indices of $Y^{\da_1\da_2}$, we will 
denote these components by $\bar{Y}$. 
We have the useful relations
\begin{equation}\label{ident}
\{Y_{0\,\da_1\da_2},Z_0^{\db_1\db_2}\} = \d_{\da_1}^{\db_1}\d_{\da_2}^{\db_2}\,,\qquad 
Y_{0\,\da_1\da_2}|E_b^0\ra_{SO(4)^2}=0\,.
\end{equation}
Using identities~\eqref{rel1}--\eqref{rel7} of appendix~\ref{appA}, the 
$SO(8)$ prefactor decomposes into the following 
$SO(4)\times SO(4)$ expressions\footnote{For the derivation of the decomposition of the 
${\mc O}(Y^6)$ term see equations~\eqref{y6_1}--\eqref{y6_5}.}
\begin{align}
K_I\wt{K}_Jw^{IJ} & = K_I\wt{K}_J\d^{IJ}\Bigl(1+\frac{1}{144}Y^4\bar{Y}^4\Bigr)\nonumber\\
&+\frac{1}{12}K_i\K_j\Bigl(\d^{ij}\bigl(Y^4+\bar{Y}^4\bigr)-3\bigl(Y^2\bar{Y}^2\bigr)^{ij}\Bigr)
\nonumber\\
&-\frac{1}{12}K_{i'}\K_{j'}\Bigl(\d^{i'j'}\bigl(Y^4+\bar{Y}^4\bigr)
+3\bigl(Y^2\bar{Y}^2\bigr)^{i'j'}\Bigr) \nonumber\\ 
&+\frac{1}{3}\bigl(K^{\da_1\a_1}\K^{\da_2\a_2}+\K^{\da_1\a_1}K^{\da_2\a_2}\bigr)
\bigl(Y^3_{\a_1\a_2}Y_{\da_1\da_2}+Y_{\a_1\a_2}Y^3_{\da_1\da_2}\bigr)\,,
\end{align}
and 
\begin{align}\label{y}
2iK_I\wt{K}_Jy^{IJ} & = K_i\K_j\Bigl(Y^{2\,ij}\bigl(1+\frac{1}{12}\bar{Y}^4\bigr)+\bar{Y}^{2\,ij}\bigl(1+\frac{1}{12}Y^4\bigr)\Bigr)
\nonumber\\
&+K_{i'}\K_{j'}\Bigl(Y^{2\,i'j'}\bigl(1-\frac{1}{12}\bar{Y}^4\bigr)+\bar{Y}^{2\,i'j'}\bigl(1-\frac{1}{12}Y^4\bigr)\bigr) \nonumber\\
& +2\bigl(K^{\da_1\a_1}\K^{\da_2\a_2}-\K^{\da_1\da_1}K^{\da_2\a_2}\bigr)\bigl(Y_{\a_1\a_2}Y_{\da_1\da_2}-\frac{1}{9}Y^3_{\a_1\a_2}
Y^3_{\da_1\da_2}\bigr)\,,
\end{align}
where we use the notation of~\cite{Pankiewicz:2003kj}, for example
\begin{equation}
K^{\da_1\a_1}=K^i{\s^i}^{\da_1\a_1}\,,\qquad {Y^2}^{ij}={Y^2}^{\a_1\b_1}\s^{ij}_{\a_1\b_1}\,,\qquad 
\bigl(Y^2\bar{Y}^2\bigr)^{ij} = Y^{2\,k(i}\bar{Y}^{2\,j)k}\,,
\end{equation}
and $Y^2_{\a_1\b_1}$ etc. are defined in appendix~\ref{appB}. 
Commuting the terms involving $\bar{Y}$ through the $Z^4$ term in 
equation~\eqref{rel_kinematical} 
using equations~\eqref{ident} and~\eqref{b1}--\eqref{b7}, one can show the 
equivalence of the two interaction Hamiltonians 
at the supergravity level
\begin{equation}
\bigl({\mc P}|V\ra\bigr)_{SO(8)\,,\,\text{Sugra}} = \bigl({\mc P}|V\ra\bigr)_{SO(4)^2\,,\,\text{Sugra}}\,.
\end{equation}
Here~\cite{Pankiewicz:2003kj}
\begin{align}
{\mc P}_{SO(4)^2} & =
\bigl(\frac{1}{2}K^{\da_1\a_1}\K^{\db_1\b_1}-\frac{\m\a}{\a'}\e^{\a_1\b_1}\e^{\da_1\db_1}\bigr)t_{\a_1\b_1}(Y)t^*_{\da_1\db_1}(Z)
\nonumber\\&
-\bigl(\frac{1}{2}K^{\da_2\a_2}\K^{\db_2\b_2}-\frac{\m\a}{\a'}\e^{\a_2\b_2}\e^{\da_2\db_2}\bigr)t_{\a_2\b_2}(Y)t^*_{\da_2\db_2}(Z)
\nonumber\\
&-K^{\da_1\a_1}\K^{\da_2\a_2}s_{\a_1\a_2}(Y)s^*_{\da_1\da_2}(Z)
-\K^{\da_1\a_1}K^{\da_2\a_2}s^*_{\a_1\a_2}(Y)s_{\da_1\da_2}(Z)\,,
\end{align}
and the spinorial quantities are 
\begin{equation}
s(Y) \equiv Y+\frac{i}{3}Y^3\,,\qquad t(Y) \equiv \e+iY^2-\frac{1}{6}Y^4\,.
\end{equation}


\section{Extension to non-zero-modes}\label{section3}

In this section, we prove that the string theory three-string interaction 
vertex constructed in the $SO(8)$ formalism in~\cite{Spradlin:2002ar,Spradlin:2002rv,Pankiewicz:2002gs,Pankiewicz:2002tg} and in the 
$SO(4)^2$  formalism in~\cite{Chu:2002eu,Chu:2002wj,Pankiewicz:2003kj} are identical.
In the $SO(8)$ formulation, the complete fermionic contribution to the kinematical part of the vertex 
is~\cite{Pankiewicz:2002gs,Spradlin:2002ar}
\begin{equation}
|E_b\ra_{SO(8)} = \exp\Bigl[\sum_{r,s=1}^3\sum_{m,n=1}^{\infty}
b_{-m(r)}^{\dag}Q^{rs}_{mn}b_{n(s)}^{\dag}-\sqrt{2}\L\sum_{r=1}^3\sum_{m=1}^{\infty}Q^r_mb_{-m(r)}^{\dag}\Bigr]
|E_b^0\ra_{SO(8)}\,.
\end{equation}
In the $SO(4)^2$ formalism the fermionic contribution to the kinematical part of the vertex is~\cite{Chu:2002wj}
\begin{align}\label{eb}
|E_b\ra_{SO(4)^2} & = \exp\Bigl[\sum_{r,s=1}^3\sum_{m,n=1}^{\infty}
\bigl(b^{\a_1\a_2\,\dag}_{-m(r)}b^{\dag}_{n(s)\,\a_1\a_2}\bar{Q}^{rs}_{mn}
-b^{\da_1\da_2\,\dag}_{-m(r)}b^{\dag}_{n(s)\,\da_1\da_2}\bar{Q}^{sr}_{nm}\bigr)
\nonumber\\
&-\sqrt{2}\L^{\a_1\a_2}\sum_{r=1}^3\sum_{m=1}^{\infty}\bar{Q}^r_mb_{-m(r)\,\a_1\a_2}^{\dag}
+\frac{\a}{\sqrt{2}}\Theta^{\da_1\da_2}\sum_{m=1}^{\infty}\bar{Q}^r_mb_{m(r)\,\da_1\da_2}^{\dag}\Bigr]
|E_b^0\ra_{SO(4)^2}\,,
\end{align}
and we have the following relations between the fermionic Neumann matrices of the two vertices
\begin{align}
Q^{rs}_{mn} & = \left(\frac{1+\Pi}{2}+\frac{1-\Pi}{2}U_{m(r)}U_{n(s)}\right)\bar{Q}^{rs}_{mn}\,,\\
Q^r_m & = \left(\frac{1+\Pi}{2}+\frac{1-\Pi}{2}(1-4\m\a K)^{-1}U_{m(r)}^{-1}\right)\bar{Q}^r_m\,.
\end{align}
The positive chirality parts of the vertices agree in both formulations. 
In what follows we 
concentrate on the contribution with negative chirality. Recall that 
$\Theta |E_b^0\ra_{SO(8)}=0$, 
($\a_3\Theta\equiv\vt_{0(1)}-\vt_{0(2)}$) and 
\begin{align}
\bar{Q}^{sr}_{nm} & = \frac{\a_rn}{\a_sm}\bar{Q}^{rs}_{mn}\,,\\
\label{idq}
\bar{Q}^{sr}_{nm}-\bigl(U_{(r)}\bar{Q}^{rs}U_{(s)}\bigr)_{mn} & = 
\bar{G}_{m(r)}\bigl(U_{(s)}\bar{G}_{(s)}\bigr)_n\,.
\end{align}
Equation~\eqref{idq} can be derived using the factorization theorem for the bosonic 
Neumann matrices~\cite{Schwarz:2002bc,Pankiewicz:2002gs}. Using these 
identities, one can show that 
the generalization of~\eqref{rel_kinematical} to include the stringy modes is
\begin{equation}\label{rel_kinematical2}
\left(\frac{2}{\a_3}\right)^4(1-4\m\a K)^{2}|E_b\ra_{SO(8)} = 
\frac{1}{12}Z^4|E_b\ra_{SO(4)^2}\,.
\end{equation}
Finally, note that 
\begin{equation}\label{fullcom}
\{Y_{\da_1\da_2},Z_{\db_1\db_2}\} = \d_{\da_1}^{\db_1}\d_{\da_2}^{\db_2}\,,\qquad 
Y_{\da_1\da_2}|E_b\ra_{SO(4)^2}=0\,.
\end{equation}
Since equations~\eqref{rel_kinematical2} and~\eqref{fullcom} are 
algebraically the same 
as~\eqref{rel_kinematical} and~\eqref{ident}, the results of section~\ref{section2} imply that
\begin{equation}
\bigl({\mc P}|V\ra\bigr)_{SO(8)} = \bigl({\mc P}|V\ra\bigr)_{SO(4)^2}\,,
\end{equation}
as conjectured in~\cite{Pankiewicz:2003kj}. 


\section{Conclusions}
In this paper, we have proved that the plane-wave light-cone superstring  
field theory Hamiltonians constructed on the states $\ket{0}_{123}$ and 
$\ket{v}_{123}$ are identical. This analysis could be easily extended to 
show the equivalence of the dynamical supercharges as well.
We have thereby resolved one of the puzzling features of the 
$SO(4)^2$ formalism, namely that it appeared not to have a smooth $\mu\to 
0$ flat space limit to 
the vertex of~\cite{Green:1983hw}. 
In fact $Z^4|E_b\ra_{SO(4)^2}\sim |E_b\ra_{SO(8)}$ and 
${\mc P}_{SO(4)^2}\bar{Y}^4\sim {\mc P}_{SO(8)}$ have well-defined limits as $\m\to0$ rather than
$|E_b\ra_{SO(4)^2}$ and ${\mc P}_{SO(4)^2}$. Moreover, since
it is known that 
$|E_b\ra_{SO(8)}$ and $|E_b\ra_{SO(4)^2}\sim\bar{Y}^4|E_b\ra_{SO(8)}$ have opposite 
$\Zop_2$ parity~\cite{Pankiewicz:2003kj,Chu:2002eu}, 
it follows that ${\mc P}_{SO(4)^2}$ and ${\mc P}_{SO(8)}$ also have opposite parity and, therefore, 
${\mc P}_{SO(4)^2}$ is odd under the $\Zop_2$. 

The existence of a smooth flat space limit, together with $\Zop_2\subset 
SO(8)$ invariance, 
suggests that the assignment of negative $\Zop_2$ parity to $\ket{v}$ (equivalently positive $\Zop_2$ parity to $\ket{0}$)
is correct: only then the plane-wave interaction Hamiltonian is invariant under $SO(4)\times SO(4)\times \Zop_2$ and the latter
is continuously connected to the $SO(8)$ symmetry of the flat space vertex.  
This suggests the uniqueness\footnote{Up to the overall normalization, which due to the absence of the $J^{-I}$ generator
can be any suitable function of the light-cone momenta.} of the 
interaction Hamiltonian at this order in the string coupling as a solution 
of the world-sheet continuity and supersymmetry algebra 
constraints.\footnote{Recently, a different solution of these conditions 
has been presented~\cite{DiVecchia:2003yp}. However, it does not have a 
smooth flat space limit and is not $\Zop_2$ invariant 
with the above parity assignment. }

The presence of apparently different interaction Hamiltonians has already 
been encountered in flat space, where two such objects were constructed. 
These had an explicit $SO(8)$ or $SU(4)$ symmetry, respectively~\cite{Green:1984fu}, 
and at first sight appear to be quite different. It is clear that 
our proof can be easily applied to show that the two expressions are, in 
fact, equivalent. Similarly for the open string interaction 
Hamiltonian in the plane-wave background, two apparently different 
expressions exist~\cite{Chandrasekhar:2003fq,Stefanski:2003zc}. Again our 
proof can be easily adapted to 
this case to show that the two are identical as operators in the three-string Hilbert space.

\section*{Acknowledgement} 
We are grateful for discussions with M.~Gaberdiel and J.~Gomis. B.~S.~is also grateful to the organizers of the Benasque workshop 
for providing a stimulating and vibrant atmosphere during the final stages 
of this project. \\ This work was supported by GIF, the German-Israeli
foundation for Scientific Research, FOM, the Dutch Foundation for
Fundamental Research on Matter and the European Community's Human
Potential Programme under contract HPRN-CT-2000-00131 in which A.~P.~is
associated to the University of Bonn and B.~S.~to the University of Utrecht. A.~P.~also acknowledges support by
the Marie Curie Research Training Site under contract HPMT-CT-2001-00296.
  

\appendix

\section{Conventions and Notation}\label{appA}

The R-R flux in the plane wave geometry breaks the $SO(8)$ symmetry of the metric into
$SO(4)\times SO(4)\times\Zop_2$. Then
\begin{equation}
{\bf 8}_v \longrightarrow ({\bf 4},{\bf 1})\oplus ({\bf 1},{\bf 4})\,,\qquad
{\bf 8}_s \longrightarrow ({\bf 2},{\bf 2}) \oplus ({\bf 2'},{\bf 2'})\,,\qquad
{\bf 8}_c \longrightarrow ({\bf 2},{\bf 2'}) \oplus ({\bf 2'},{\bf 2})\,,
\end{equation}
where ${\bf 2}$ and ${\bf 2'}$ are the inequivalent Weyl representations of $SO(4)$.
We decompose $\g^I_{a\dot{a}}$ and $\g^I_{\dot{a}a}$ into $SO(4)\times SO(4)$ as follows
\begin{align}
\g^i_{a\dot{a}} & =
\begin{pmatrix}
0 & \s^i_{\a_1\db_1}\d_{\a_2}^{\b_2} \\ {\s^i}^{\da_1\b_1}\d^{\da_2}_{\db_2} & 0
\end{pmatrix}\,,\qquad
\g^i_{\dot{a}a} =
\begin{pmatrix}
0 & \s^i_{\a_1\db_1}\d^{\da_2}_{\db_2} \\ {\s^i}^{\da_1\b_1}\d_{\a_2}^{\b_2} & 0
\end{pmatrix}\,,\\
\g^{i'}_{a\dot{a}} & =
\begin{pmatrix}
-\d_{\a_1}^{\b_1}\s^{i'}_{\a_2\db_2} & 0 \\ 0 & \d^{\da_1}_{\db_1}{\s^{i'}}^{\da_2\b_2}
\end{pmatrix}\,,\qquad
\g^{i'}_{\dot{a}a} =
\begin{pmatrix}
-\d_{\a_1}^{\b_1}{\s^{i'}}^{\da_2\b_2} & 0 \\ 0 & \d^{\da_1}_{\db_1}\s^{i'}_{\a_2\db_2}
\end{pmatrix}\,.
\end{align}
Here, the $\s$-matrices consist of the usual Pauli-matrices, together with 
the 2d unit matrix
\begin{equation}
\s^i_{\a\da}=\bigl(i\t^1,i\t^2,i\t^3,-1\bigr)_{\a\da}
\end{equation}
and we raise and lower spinor indices with the two-dimensional Levi-Civita symbols, e.g.
\begin{equation}
\s^i_{\a\da} = \e_{\a\b}\e_{\da\db}{\s^i}^{\db\b}
\equiv \e_{\a\b}{\s^i}^{\b}_{\da} \equiv \e_{\da\db}{\s^i}^{\db}_{\a}\,.
\end{equation}
The $\s$-matrices obey the relations
\begin{equation}
\s^i_{\a\da}{\s^j}^{\da\b}+\s^j_{\a\da}{\s^i}^{\da\b}
=2\d^{ij}\d_{\a}^{\b}\,,\qquad
{\s^i}^{\da\a}\s^j_{\a\db}+{\s^j}^{\da\a}\s^i_{\a\db}=
2\d^{ij}\d^{\da}_{\db}\,.
\end{equation}
In particular, in this basis
\begin{equation}
\Pi_{ab} =
\begin{pmatrix} \bigl(\s^1\s^2\s^3\s^4\bigr)_{\a_1}^{\b_1}\d_{\a_2}^{\b_2} & 0 \\
0 & \bigl(\s^1\s^2\s^3\s^4\bigr)^{\da_1}_{\db_1}\d^{\da_2}_{\db_2}
\end{pmatrix} =
\begin{pmatrix}
\d_{\a_1}^{\b_1}\d_{\a_2}^{\b_2} & 0 \\ 0 & -\d^{\da_1}_{\db_1}\d^{\da_2}_{\db_2}
\end{pmatrix}\,,
\end{equation}
and $(1\pm\Pi)/2$ projects onto $({\bf 2},{\bf 2})$ and
$({\bf 2'},{\bf 2'})$, respectively. The following identities are used 
throught the paper
\begin{align}
\label{rel1}
\e_{\a\b}\e^{\g\d} & = \d_{\a}^{\d}\d_{\b}^{\g}-\d_{\a}^{\g}\d_{\b}^{\d}\,,\\
\s^i_{\a\db}{\s^j}^{\db}_{\b} & = -\d^{ij}\e_{\a\b}+\s^{ij}_{\a\b}\,,\qquad
(\s^{ij}_{\a\b}\equiv \s^{[i}_{\a\da}{\s^{j]}}^{\da}_{\b}=\s^{ij}_{\b\a})\\
\s^i_{\a\da}{\s^j}^{\a}_{\db} & = -\d^{ij}\e_{\da\db}+\s^{ij}_{\da\db}\,,
\qquad (\s^{ij}_{\da\db}\equiv \s^{[i}_{\a\da}{\s^{j]}}^{\a}_{\db}=\s^{ij}_{\db\da})\\
\s^k_{\a\da}\s^{k}_{\b\db} & = 2\e_{\a\b}\e_{\da\db}\,,\\
\s^{ik}_{\a\b}\s^k_{\g\dd} & = \e_{\a\g}\s^i_{\b\dd}+\e_{\b\g}\s^i_{\g\dd}\,,\\
\s^{ik}_{\a\b}\s^{jk}_{\g\d} & = \d^{ij}(\e_{\a\g}\e_{\b\d}+\e_{\a\d}\e_{\b\g})-\frac{1}{2}
\bigl(\s^{ij}_{\a\g}\e_{\b\d}+\s^{ij}_{\b\d}\e_{\a\g}+\s^{ij}_{\a\d}\e_{\b\g}+\s^{ij}_{\b\g}\e_{\a\d}\bigr)
\,,\\
\s^{kl}_{\a\b}\s^{kl}_{\g\d} & = 4(\e_{\a\g}\e_{\b\d}+\e_{\a\d}\e_{\b\g})\,,\\
\s^{kl}_{\a\b}\s^{kl}_{\dg\dd} & = 0\,,\\
\label{rel7}
2\s^i_{\a\da}\s^{j}_{\b\db} & = \d^{ij}\e_{\a\b}\e_{\da\db}
+\s^{k(i}_{\a_1\b_1}\s^{j)k}_{\da_1\db_1}
-\e_{\a\b}\s^{ij}_{\da\db}-\s^{ij}_{\a\b}\e_{\da\db}\,.
\end{align}

\section{Useful relations}\label{appB}

We define the following quantities, which are quadratic in $Y$ and 
symmetric in spinor indices
\begin{equation}\label{y2}
Y^2_{\a_1\b_1} \equiv Y_{\a_1\a_2}Y^{\a_2}_{\b_1}\,,\qquad Y^2_{\a_2\b_2} \equiv Y_{\a_1\a_2}Y^{\a_1}_{\b_2}\,,
\end{equation}
cubic in $Y$
\begin{equation}\label{y3}
Y^3_{\a_1\b_2} \equiv Y^2_{\a_1\b_1}Y^{\b_1}_{\b_2}=-Y^2_{\b_2\a_2}Y^{\a_2}_{\a_1}\,,
\end{equation}
and, finally, quartic in $Y$ and antisymmetric in spinor indices
\begin{equation}
Y^4_{\a_1\b_1} \equiv Y^2_{\a_1\g_1}{Y^2}^{\g_1}_{\b_1}=-\frac{1}{2}\e_{\a_1\b_1}Y^4\,,\qquad
Y^4_{\a_2\b_2} \equiv Y^2_{\a_2\g_2}{Y^2}^{\g_2}_{\b_2}=\frac{1}{2}\e_{\a_2\b_2}Y^4\,,
\end{equation}
where
\begin{equation}\label{y4}
Y^4 \equiv Y^2_{\a_1\b_1}{Y^2}^{\a_1\b_1}=-Y^2_{\a_2\b_2}{Y^2}^{\a_2\b_2}\,.
\end{equation}
These multi-linears in $Y$ satisfy
\begin{align}
\label{id1}
Y_{\a_1\a_2}Y_{\b_1\b_2} & = -\frac{1}{2}
\bigl(\e_{\a_1\b_1}Y^2_{\a_2\b_2}+\e_{\a_2\b_2}Y^2_{\a_1\b_1}\bigr)\,,\\
\label{id2}
Y_{\a_1\a_2}Y^2_{\b_2\g_2} & = -\frac{1}{3}\bigl(
\e_{\a_2\g_2}Y^3_{\a_1\b_2}+\e_{\a_2\b_2}Y^3_{\a_1\g_2}\bigr)\,,\\
\label{id3}
Y_{\a_1\a_2}Y^2_{\b_1\g_1} & = \frac{1}{3}\bigl(
\e_{\a_1\b_1}Y^3_{\g_1\a_2}+\e_{\a_1\g_1}Y^3_{\a_1\a_2}\bigr)\,,\\
\label{id4}
Y^3_{\b_1\g_2}Y_{\a_1\d_2} & = \frac{1}{4}\e_{\b_1\a_1}\e_{\g_2\d_2}Y^4\,.
\end{align}
Analogous relations hold for $Z$.

To derive equations~\eqref{rel_kinematical} and~\eqref{rel_kinematical2} we need the following (anti)commutators
\begin{align}
\label{b1}
[Y_{\da_1\da_2},Z^2_{\db_1\dg_1}] & = \e_{\da_1\db_1}Z_{\dg_1\da_2}+\e_{\da_1\dg_1}Z_{\db_1\da_2}\,,\\
[Y_{\da_1\da_2},Z^2_{\db_2\dg_2}] & = \e_{\da_2\db_2}Z_{\da_1\dg_2}+\e_{\da_2\dg_2}Z_{\da_1\db_2}\,,\\
\{Y_{\da_1\da_2},Z^3_{\db_1\db_2}\} & = -3Z_{\da_1\db_2}Z_{\db_1\da_2}\,,\\
[Y_{\da_1\da_2},Z^4] & = -4Z^3_{\da_1\da_2}\,,\\
[Y^2_{\da_1\db_1},Z^4]|E_b\ra_{SO(4)^2} & = -12Z^2_{\da_1\db_1}|E_b\ra_{SO(4)^2}\,,\\
[Y^2_{\da_2\db_2},Z^4]|E_b\ra_{SO(4)^2} & = 12Z^2_{\da_2\db_2}|E_b\ra_{SO(4)^2}\,,\\
[Y^3_{\da_1\da_2},Z^4]|E_b\ra_{SO(4)^2} & = -36Z_{\da_1\da_2}|E_b\ra_{SO(4)^2}\,,\\
\label{b7}
[\bar{Y}^4,Z^4]|E_b\ra_{SO(4)^2} & = 144|E_b\ra_{SO(4)^2}\,,
\end{align}
Finally, to rewrite the ${\mc O}(Y^6)$ term in the $SO(8)$ prefactor in a manifestly $SO(4)\times SO(4)$ invariant form, it is useful 
to employ the identity~\cite{Green:1983hw}
\begin{equation}\label{y6_1} 
-\frac{1}{6!}\g^{IJ}_{ab}{\e^{ab}}_{cdefgh}Y^cY^dY^eY^fY^gY^h = \int d^8\L\,\g^{IJ}_{ab}\L^a\L^be^{-Y\cdot\L}\,,
\end{equation}
and 
\begin{align}
\label{y6_2}
\int\prod_{\a_1\a_2}d\L_{\a_1\a_2}\,e^{-Y_{\g_1\g_2}\L^{\g_1\g_2}} & = -\frac{1}{12}Y^4\,,\\
\int\prod_{\a_1\a_2}d\L_{\a_1\a_2}\,\L_{\b_1\b_2}e^{-Y_{\g_1\g_2}\L^{\g_1\g_2}} & = -\frac{1}{3}Y^3_{\b_1\b_2}\,,\\
\int\prod_{\a_1\a_2}d\L_{\a_1\a_2}\,\L^2_{\b_1\d_1}e^{-Y_{\g_1\g_2}\L^{\g_1\g_2}} & = Y^2_{\b_1\d_1}\,,\\
\label{y6_5}
\int\prod_{\a_1\a_2}d\L_{\a_1\a_2}\,\L^2_{\b_2\d_2}e^{-Y_{\g_1\g_2}\L^{\g_1\g_2}} & = -Y^2_{\b_2\d_2}\,.
\end{align}

\providecommand{\href}[2]{#2}\begingroup\raggedright

\endgroup

\end{document}